\documentstyle[12pt,epsfig]{article}                                                         
                                                         
\newlength{\dinwidth}                                                         
\newlength{\dinmargin}                                                         
\def\lapproxeq{\lower .7ex\hbox{$\;\stackrel{\textstyle                                                         
<}{\sim}\;$}}                                                         
\def\gapproxeq{\lower .7ex\hbox{$\;\stackrel{\textstyle                                                         
>}{\sim}\;$}}                                                         
\def\be{\begin{equation}}                                                         
\def\ee{\end{equation}}                                                         
\def\bea{\begin{eqnarray}}                                                         
\def\eea{\end{eqnarray}}                                                         
                                                         
\def\funp{{I\!\!P}}                         
                                                         
\begin{document}                                                         
\titlepage                                                         
\begin{flushright}                                                         
IPPP/00/01 \\   
DTP/00/58 \\                                                         
16 October 2000 \\                                                         
\end{flushright}                                                         
                                                         
\vspace*{2cm}                                                         
                                                         
\begin{center}                                                         
{\Large \bf Luminosity monitors at the LHC}                                                         
                                                         
\vspace*{1cm}                                                         
V.A. Khoze$^a$, A.D. Martin$^a$, R. Orava$^b$ and M.G. Ryskin$^{a,c}$ \\                                                         
                                                        
\vspace*{0.5cm}                                                         
$^a$ Department of Physics and Institute for Particle Physics Phenomenology, University of   
Durham, Durham, DH1 3LE \\      
$^b$ Department of Physics, University of Helsinki, and Helsinki Institute of Physics,   
Finland \\                                                      
$^c$ Petersburg Nuclear Physics Institute, Gatchina, St.~Petersburg, 188300, Russia                 
\end{center}                                                         
                                                         
\vspace*{2cm}                                                         
                                                         
\begin{abstract}                                                         
We study the theoretical accuracy of various methods that have been proposed to measure the    
luminosity of the LHC $pp$ collider, as well as for Run II of the Tevatron $p\bar{p}$    
collider. In particular we consider methods based on (i) the total and forward elastic data, (ii)    
lepton-pair production and (iii) $W$ and $Z$ production.   
\end{abstract}                                               
          
\newpage                
\section{Introduction}

The Large Hadron Collider (LHC) being constructed at CERN, will generate proton-proton    
collisions with total c.m. energy of 14 TeV with a design luminosity $\mathcal{L}$ $=   
10^{34}$ cm$^{-2}$ s$^{-1}$.  The experiments at this new facility will have a high   
potential to discover New Physics and to make various precision measurements, see,    
for example \cite{R1,R2}. Both the general purpose $pp$ experiments, ATLAS \cite{R3}    
and CMS \cite{R4}, will provide high statistics data samples, and the accuracy of the   
precision measurements will be limited by systematic effects and, in many cases, by the   
uncertainty in the measurement of the luminosity $\mathcal{L}$. For example, precision   
measurements in the Higgs sector of typical accuracy of about 7\%, for a wide range of   
possible Higgs mass, require the uncertainty in the luminosity to be $\leq 5\%$ \cite{R2}.   
Both LHC experiments are aiming at measuring the luminosity to 5\%.  Recall also that the   
enhancement in luminosity that will be achieved in Run II of the Fermilab Tevatron   
$p\bar{p}$ - collider will herald a new era of precision studies \cite{R5,R6,R7}.   
   
An obvious requirement for the success of these precision measurements is that the    
uncertainties in the theoretical calculations of the cross sections for the basic processes, which    
are to be used to determine the luminosity, should match the desired experimental accuracies.   
   
In general, there are two possibilities to determine the luminosity -- either (i) to measure a    
pair of cross sections which are connected quadratically with each other, or (ii) to measure a    
cross section whose value is well known or which may be calculated with good accuracy. The    
well-known example of the first possibility is the measurement of the total    
$\sigma_{\rm{tot}}$ and differential forward elastic cross sections which are related by the    
optical theorem; see, for example, \cite{R5,R8} for recent experimental discussions. This    
method is discussed further in Section 2.   
   
Two types of processes stand out as examples of the second possibility to measure the    
luminosity. First there is exclusive lepton-pair production via photon-photon fusion   
\be   
\label{eq:G1}   
\stackrel{}{p} \stackrel{(-)}{p} \rightarrow \ \stackrel{}{p} + \ l^+l^- \ + \stackrel{(-)}{p}   
\ee   
where $l = e$ or $\mu$. To the best of our knowledge this proposal originated in \cite{R9}.   
A luminometer for the LHC based on measuring forward $e^+e^-$ pairs (of invariant mass    
$M_{ee}< 10-20$ MeV and transverse momentum $p_t(ee) < 10-20$ MeV) was proposed in    
\cite{R10}, while Ref. \cite{R11} concerns the central production of $\mu^+\mu^-$ pairs    
(with $\left<M_{\mu\mu}\right> \ \sim$ 20 GeV and $p_t(\mu\mu) \sim$ 10--50~MeV).    
Lepton-pair production is the subject of Section 3.   
   
Nowadays attention has also focussed on $W$ and $Z$ production as a possible luminosity    
monitor, both for Run II at the Tevatron and for the LHC see, for example, \cite{R12}. The    
reason is that the signal is clean, and the production cross sections are large and can now be   
calculated with considerable theoretical accuracy. We discuss this possibility  further in   
Section 4.   
   
In principle, we may {\it monitor} the luminosity using {\it any} process, with a significant     
cross section, which is straightforward to detect.  For example, it could be single- (or two-)     
pion inclusive production in some rapidity and $p_t$ domain (for example $p_t = 5-10~{\rm   
GeV}; |\eta| < 1$) or inclusive $\mu^+ \mu^-$ production in a well-defined kinematic   
domain, etc.  In this way we may control the relative luminosity and then calibrate the \lq\lq   
monitor\rq\rq\ by comparing the number of events detected for the \lq\lq monitor\rq\rq\   
reaction with the number of events observed for a process with a cross section which is   
already known, or which may be calculated with sufficient accuracy.     
   
For some applications better accuracy can be achieved by measuring the parton-parton    
luminosity.  A discussion is given in Section~5.  Finally, Section~6 contains our conclusions.   
    
\section{Elastic scattering}    
    
First we discuss the classic method to measure the luminosity, that is using the observed total     
and forward elastic $pp$ (or $p\bar{p}$) event rates.  Neglecting Coulomb    
effects\footnote{Strictly speaking,  we have to use Coulomb wave functions, rather than plane    
waves, for the in- and out-states for elastic scattering between electrically charged protons.     
To account for this effect, the strong interaction amplitude $A_{\rm el}$ should be multiplied    
by the well-known Bethe phase \cite{BWY} $(A_{\rm el} \rightarrow \exp (i \delta_B)    
A_{\rm el})$.  The uncertainty in Bethe phase $\delta_B$ due to the non-point-like structure    
(electric charge distribution) of the proton is about $\alpha = 1/137$.  It leads to an    
uncertainty of about $\pm 0.01$ in the Re/Im ratio, and hence to less than a 0.1\% correction    
for the imaginary part of the strong amplitude $A_{\rm el}$.  A recent discussion of the  
Coulomb phase for $t \neq 0$ can be found in \cite{KT}.}, the elastic cross section is   
given by    
\be    
\label{eq:2a}    
\left . \frac{d \sigma_{\rm el}}{dt} \right |_{t = 0} \; = \; \frac{\sigma_{\rm tot}^2}{16 \pi} \:     
(1 + \rho^2).    
\ee    
The ratio $\rho$ of the real to the imaginary parts of the forward amplitude is small at     
Tevatron-LHC energies and can be estimated via dispersion relation techniques    
\be    
\label{eq:2b}    
\rho \; \simeq \; \frac{\pi}{2} \: \frac{\partial \ln \sigma_{\rm tot} (s)}{\partial \ln s}.    
\ee    
For example, at the LHC energy, $\sqrt{s} = 14~{\rm TeV}$, it is predicted to be $\rho =     
0.10-0.12$; for a recent estimate, see, for example, Ref.~\cite{KMR}.  Even the largest    
conceivable uncertainty in $\rho,$ that is $\Delta \rho = \pm 0.02$, leads only to an    
uncertainty of less than 0.5\% in $d\sigma_{\rm el}/dt$.  Thus if we measure the number of    
events corresponding to the elastic scattering and to the total cross section, then we may    
determine both the luminosity ${\cal L}$ and $\sigma_{\rm tot}$ (since $N_{\rm el} \propto    
\sigma_t^2 {\cal L}$, whereas $N_{\rm tot} \propto \sigma_t {\cal L}$).  The main problem    
in that it is extremely difficult to make measurements at the LHC in the very forward region,    
and so it is necessary to extrapolate elastic data from, say, $|t| \gapproxeq 0.01~{\rm    
GeV}^2$ to $t = 0$.    
    
On the other hand it was found at the ISR that the observed \lq local\rq\ slope    
\be    
\label{eq:2c}    
B (t) \; \equiv \; \frac{d \ln (d \sigma_{\rm el}/dt)}{dt}    
\ee    
depends on $t$.  In particular at $\sqrt{s} = 62~{\rm GeV}$ \cite{ISR}    
\be    
\label{eq:2d}    
\Delta B \; = \; B (0) - B (|t| = 0.2~{\rm GeV}^2) \; \simeq \; 2~{\rm GeV}^{-2}.    
\ee    
Note that if this $t$ dependence were to be solely due to a non-linear Pomeron trajectory    
\cite{AG}, then the difference (\ref{eq:2d}) would increase as the logarithm of the energy    
$(\ln s)$; giving, for example, $\Delta B \simeq 4.8~{\rm GeV}^{-2}$ at the    
LHC energy.  The $t$-dependence of the local slope $B (t)$ has recently been determined    
\cite{KMR} using a model which incorporates all the main features of high energy soft    
diffraction.  That is the model embodies    
\begin{itemize}    
\item[(a)] the {\it pion-loop} contribution to the Pomeron pole (which is the main source of     
the non-linearity of the Pomeron trajectory $\alpha_\funp (t)$),    
\item[(b)] a {\it two-channel} eikonal to include the Pomeron cuts which are generated by     
elastic and quasi-elastic (with $N^*$ intermediate states) $s$-channel unitarity,    
\item[(c)] the effects of high-mass {\it diffractive dissociation}.    
\end{itemize}    
    
The parameters of the model are $\alpha (0)$ and $\alpha^\prime$ of the {\it bare} Pomeron     
trajectory, two parameters describing the elastic proton-Pomeron vertex, and the     
triple-Pomeron coupling and its slope \cite{KMR}.  The values of the parameters were tuned     
to describe the observed $pp$ (or $p\bar{p}$) elastic differential cross sections throughout    
the ISR-Tevatron range, restricting the description to the forward region $|t| < 0.5~{\rm    
GeV}^2$.  Note that there are two effects, leading to a $t$ dependence of $B$, which act in    
opposite directions.  First, the non-linear pion-loop contributions \cite{AG} to the Pomeron    
trajectory, $\Delta \alpha_\funp (t)$, lead to a contribution $\Delta B_\funp (t) = \Delta    
\alpha_\funp (t) \ln s$ in single Pomeron exchange, which increases as $|t| \rightarrow 0$.  On    
the other hand the absorptive (rescattering) corrections, associated with eikonalization, lead to    
a dip in $d \sigma_{\rm el}/dt$ (whose position moves to smaller $|t|$ as the collider energy,    
$\sqrt{s}$, increases), with the result that the local slope $B (t)$ grows as $-t$ approaches the    
position of the diffractive minimum; that is $B (t)$ decreases as $|t| \rightarrow 0$.     
Fortunately, just at the LHC energy, these two effects almost compensate each other.      
    
So far it has proved impossible to sum up in a consistent way all the multi-Pomeron    
diagrams.  For this     
reason two versions of the model were studied in Ref.~\cite{KMR} with {\it maximal} and     
{\it minimal} contributions from high-mass diffractive dissociation.  After tuning the values     
of the parameters to describe the $\sigma_{\rm tot}$ and the forward $d \sigma_{\rm el}/dt$     
data, the two versions of the model predict different values of the slope at $t = 0, B (0)$.  The     
maximal choice gives a larger value $B (0) = 21.9~{\rm GeV}^{-2}$, while for the minimal     
case we obtain $B (0) = 20.3~{\rm GeV}^{-2}$.  However in both cases the variation of $B     
(t)$ satisfies $\Delta B < 0.2~{\rm GeV}^{-2}$ for $|t| < 0.1~{\rm GeV}^2$, or $\Delta B <     
0.1~{\rm GeV}^{-2}$ for the more restricted interval $0 < |t| < 0.02~{\rm GeV}^2$.  It    
means     
that we may neglect such a small variation of $B (t)$ with $t$ and safely extrapolate the    
observed     
cross section $d \sigma_{\rm el}/dt$ down to $t = 0$, using the slope measured, say, in the     
interval $0.01 < |t| < 0.05~{\rm GeV}^2$.  The error, due to the variation of $B$ with $t$, is     
expected to be less than $\Delta B \cdot |t| \lapproxeq 0.2\%$.    
    
In the very forward region, $|t| \sim 0.01-0.02~{\rm GeV}^2$, it may only be     
possible to measure the elastic cross section in an LHC run with a rather low luminosity.  For     
the high luminosity runs one would then have to use another monitor.  One possibility is to     
choose a single particle inclusive process, with a significant cross section, and \lq\lq     
calibrate\rq\rq\ it in the low luminosity run by comparing with the total and elastic cross     
sections discussed above.  It would be even better if we were able to calibrate several     
different reactions.  Then to use, as a luminosity monitor, the reaction with the closest     
topology (or kinematic configuration) to the process that we are to study in the high     
luminosity run.    
    
\section{Lepton pair production as a luminometer}    
    
At first sight any QED process, with sufficient event rate, may be used as a luminosity     
monitor.  Up to ISR energies the luminosity was measured via Coulomb elastic $pp$     
scattering, where at very small $|t|$ the influence of the dominant single photon exchange     
contribution is evident.  Unfortunately, in the LHC environment, the Coulomb interference    
region, $|t|  \lapproxeq (1-3) \times 10^{-4}~{\rm GeV}^2$ is not likely to be experimentally    
accessible.    
    
Other possibilities to consider are $\Delta$ isobar Coulomb excitation, ($pp \rightarrow     
\Delta^+ p$), or $\pi^0$ and $\eta$ production, which are all mediated by photon exchange,    
as shown in Figs.~1(a) and 2(a) respectively.  Surprisingly, the electromagnetic widths    
$\Delta^+ \rightarrow p\gamma, \pi^0 \rightarrow \gamma\gamma$ and $\eta \rightarrow    
\gamma\gamma$ are known only to 7--10\% accuracy \cite{PDG}.  Moreover, strong    
interaction effects in the initial and final states give non-negligible corrections, see Figs.~1(b)    
and 2(b).    
    
Lepton pair production looks much more promising as a luminosity monitor.  The Born     
amplitude of Fig.~3(a) may be calculated within pure QED (see, for example, the reviews in    
\cite{BB}), and there are no strong interactions involving the leptons in the final state.  The    
only question is the size of the absorptive corrections arising from inelastic proton-proton    
rescattering, sketched in Fig.~3(b).      
Fortunately the rescattering correction is suppressed for two reasons.  First, the main part of     
the Born cross section (Fig.~3(a)) comes from the peripheral region with large impact     
parameter $b_t$, where the strong amplitude $A (s, b_t)$ is small.  Second, even in the small     
$b_t$ domain, the rescattering correction is greatly suppressed due to the angular integration.    
    
Before we discuss the last point, we note that, in practice, it is difficult to exclude    
contributions coming from the reactions   
\bea      
\label{eq:F1}    
pp \rightarrow X + l^+ l^- + p \nonumber\\   
pp \rightarrow X + l^+ l^- + Y    
\eea   
where $X$ and $Y$ are baryon excitations, that is $N^*$ or $\Delta$ isobars. Of course the    
matrix elements of the corresponding processes (such as Fig.~3(c)) can, in principle, be    
determined from photoproduction and deep inelastic data, and may be taken into account.    
However the matrix elements of Fig.~3(c) are not known to sufficient accuracy and it is better    
to suppress the contributions of (\ref{eq:F1}) by experimental cuts. The procedure is as    
follows. Recall that, due to gauge invariance, inelastic vertices of the type $p \rightarrow X +    
\gamma$ vanish like \cite{BB}   
\be     
\label{eq:F2}    
V(p \rightarrow X\gamma) \ \propto \ q_{1t}   
\ee   
as the photon transverse momentum $q_{1t} \rightarrow 0$. Unfortunately it is difficult to    
measure a leading proton with very small transverse momentum, that is $q_t \lapproxeq    
10-30~{\rm MeV}$.  So to take advantage of the behaviour of (\ref{eq:F2}), it was proposed    
\cite{R11} to select events with very small transverse momentum of the lepton pair   
\be   
\label{eq:F3}    
p_t \equiv \left | \ \mbox{\boldmath $q$}_{l^+t} + \mbox{\boldmath $q$}_{l^-t} \ \right | <    
10-30 \rm{MeV},   
\ee   
in order to suppress $N^*$ and $\Delta$ production.  In such a case the integral over the    
transverse momenta of the photons in the Born cross section of Fig.~3(a) takes the form   
\be   
\label{eq:F4}   
\int \: \frac{dq_{1t}^2 \: dq_{2t}^2 \: q_{1t}^2 \: q_{2t}^2}{(q_{1t}^2 \: + \: t_1)^2   
(q_{2t}^2 \: + \: t_2)^2} \: \delta^{(2)} (\mbox{\boldmath $p$}_t - \mbox{\boldmath   
$q$}_{1t} - \mbox{\boldmath $q$}_{2t}) \: d^2 p_t,   
\ee   
where the dominant contributions come from the regions $q_{it} \lapproxeq p_t$.  The values    
of the longitudinal components are   
\be   
\label{eq:F5}   
t_i \equiv \left | t_{i,min} \right | = x^2_i m^2_p ,   
\ee   
where $x_i$ are the fractions of the momenta of the incoming protons carried by the $l^+ l^-$    
pair,   
and $m_p$ is the mass of the proton.   
   
\subsection{$\mu^+ \mu^-$ production}   
   
To identify muons (and to separate them from $\pi^\pm$ mesons) it is necessary to observe    
charged particles after they have transversed a thick (iron) absorber.  It means that we    
consider only muons which have a rather large transverse energy $E_t \gapproxeq 5$ GeV.     
However, it is still possible to select events where the sum of their transverse momenta is    
small, $p_t < 30$ MeV. In this case    
\be   
\label{eq:F6}   
x_i \simeq (2-3) E_t/\sqrt{s} \simeq 10^{-3}, \ \ \ \ \ t_i \simeq 10^{-6} \rm{GeV}^2 ,   
\ee   
at the LHC energy. In such a configuration the main contribution to the integral (\ref{eq:F4})   
comes from the domains $q_{1t} \approx p_t,~t_2 \ll q_{2t}^2 \ll p_t^2$ and $q_{2t}    
\approx p_t,~t_1 \ll q_{1t}^2 \ll p_t^2$.  Performing the $q_{it}^2$ integrations in these two    
domains gives $\ln (p_t^2/t_2)$ and $\ln (p_t^2/t_1)$ respectively, and so the Born cross    
section behaves as   
\be   
\label{eq:F7}   
\frac {d \hat{\sigma}}{dp^2_t} \ \propto \ \frac {1}{p^2_t} \ {\rm ln} \left (    
\frac{p^4_t}{t_1t_2}    
\right )   
\ee   
  
In addition to the use of the small $p_t$ cut in order to separate the elastic process  
(\ref{eq:G1}) from the excitation process (\ref{eq:F1}), we can also exploit the different  
kinematics of the processes.  For example in \cite{R11} it was proposed to fit the observed  
distribution in the muon acoplanarity angle $\phi$ in order to isolate the elastic mechanism  
via its prominent peak at $\phi = 0$.  
   
\subsection{Absorptive corrections to $l^+l^-$ production}   
   
To determine the effect of the absorptive, or re-scattering, correction, we calculate diagram    
3(b) with an extra loop integration over the momentum $Q$ transferred via the strong    
interaction amplitude shown by the \lq\lq blob". The relative size of the correction, to    
the amplitude for lepton-pair production of Fig.~3(a), is   
\be    
\label{eq:F8}   
\delta \; = \; \frac{(\sigma_{\rm inel}/8\pi^2) \int d^2Q_t d^2q_{1t} d^2p_t \hat{\sigma}    
(q_1,q_2;q_1+Q,q_2+Q) \: F (Q)}{\int d^2q_{1t} d^2p_t \hat{\sigma} (q_1,q_2;q_1,q_2) \: F   
(0)},   
\ee   
where $F (Q)$ represents the collective effect of the appropriate proton electromagnetic form  
factors $F_N$ and the $Q^2$ dependence of the strong interaction amplitude  
\be  
\label{eq:F8b}  
F (Q) \; = \; F_N (q_1^2) \: F_N (q_2^2) \: F_N ((Q + q_1)^2) \: F_N ((Q + q_2)^2) \: e^{b   
Q^2/2}.  
\ee  
$b$ is the slope of the $pp$ elastic cross section, that is $d \sigma_{\rm el}/dt \propto \exp  
(bt)$.  We can interpret the correction $\delta$ in terms of $S = 1 - \delta$, where $S$ is the  
\lq\lq   
survival probability amplitude\rq\rq\ that the secondary hadrons produced in the soft   
rescattering do not accompany $\ell^+ \ell^-$ production.  Of  course this effect is due to the   
inelastic strong interaction only.  Consider the hypothetical case with pure elastic $pp$   
rescattering (at fixed impact parameter $b_t$, that is for fixed partial wave $\ell = b_t   
\sqrt{s}/2$).  Then elastic rescattering only changes the phase of the QED matrix element,   
${\cal M} \rightarrow {\cal M} \exp (2 i \phi_\ell)$, and does not alter the QED cross section  
$\sigma = |{\cal M}|^2$.  Thus we have to account only for the inelasticity of the strong  
interaction, that is $\delta \propto \sigma_{\rm inel}$ as in (\ref{eq:F8})\footnote{An  
alternative, and more formal, explanation of $\delta \propto \sigma_{\rm inel}$ is that we  
have to sum up multiple $pp$ rescattering.  On resumming all the eikonal graphs, it turns out  
that one obtains (\ref{eq:F8}) with $\sigma_{\rm tot} - \sigma_{\rm el} = \sigma_{\rm  
inel}$.}.  Throughout this paper we do not discuss the effects of \lq\lq pile-up\rq\rq\ events.   
Thus there will be an additional factor $(W < 1)$ which represents the probability not to  
include events with extra secondaries coming from an almost simultaneous interaction of  
another pair of protons.  Experimentally the depletion due to such pile-up events may be  
overcome, for example, by cleanly observing the vertex of $\mu^+ \mu^-$ production. 
  
The cross section $\hat{\sigma}$ in (\ref{eq:F8}) plays the role of the cross section for the   
QED subprocess of $\gamma \gamma$ scattering through the lepton box. For the absorptive   
process, the photons have momenta $q_1 + Q$ and $q_2 + Q$ in the \lq\lq left" amplitude   
$A$ shown in Fig.~3(b), and $q_1, q_2$ in the \lq\lq right" amplitude $A^*$, which is not    
shown.  Here we have assigned the absorptive effect to $A$. Actually the numerator of  
(\ref{eq:F8}), and the symbolic\footnote{Indeed Fig.~3(b) is truly symbolic and must not be  
viewed literally.  The problem is that the strong interaction is not mediated by a point-like  
object, but must rather be viewed as a multipheral or gluon ladder (Pomeron) exchanged  
between the protons.  It turns out that the dominant contribution comes from four different  
configurations, corresponding to the photons being emitted either before or inside the  
Pomeron ladder.  Due to the conservation of the electromagnetic current, the sum of all four  
contributions is embodied in (\ref{eq:F8}) --- the only qualification is that when the photon is  
emitted inside the ladder, the form factor $F_N$ may have different behaviour at large $Q_t$.   
In our case, when $Q_t$ is very small, this effect is negligible.} diagram 3(b), summarize a  
set of Feynman graphs which describe the effects of the strong interaction between the  
protons in the amplitude $A$.  Then we have to add the equivalent set of diagrams in which  
the rescattering effects occur in the amplitude $A^*$.  The total correction to (\ref{eq:F7}) is  
therefore $2\delta d\hat{\sigma}/dp_t^2$.  The rescattering correction is due to the imaginary  
part of the strong amplitude only.  The effects coming from the real part in $A$ cancel with  
those in $A^*$.   
   
At first sight the largest absorptive effects, (\ref{eq:F8}), appear to come from the largest    
values of $Q_t \lapproxeq 1/R_p$, where $R_p$ is the proton radius; higher values of $Q_t$    
are cut-off by the proton form factor. However when $Q_t \gg q_{1t},q_{2t}$ an interesting    
suppression occurs. In this domain $\gamma \gamma \rightarrow l^+l^-$ occurs dominantly    
in the $J_z = 0$ two-photon state. The projection of the orbital angular momentum on the    
longitudinal $(z)$ axis is clearly zero, while the spin (which originates from the photon    
polarisations) is described by the tensor $Q_{\mu t} Q_{\nu t}$. After the azimuthal    
integration, the tensor takes the form $\frac{1}{2} Q^2 \delta^{\bot}_{\mu\nu}$, leading to    
$J_z = 0$.  If we neglect the mass of the lepton, $m_l \ll E_t$, and higher order QED    
corrections, then the Born amplitude, with $J_z = 0$, vanishes\footnote{An analogous    
cancellation, which occurs on integration over the azimuthal angle, was observed in QCD for    
light quark-pair production $pp \rightarrow p + q\bar{q} + p$ by Pumplin \cite{PUMP}, see    
also \cite{BC,KMR2}.}.  This is a well known result, see, for example, \cite{KAI}.  As    
shown in Ref.~\cite{BKSO}, the physical origin of the suppression is related to the symmetry    
properties of the $J_z = 0$ Born amplitude.   
   
The \lq\lq cross section\rq\rq\ $\hat{\sigma}(q_1,q_2;q_1+Q,q_2+Q)$ for our QED    
subprocess depends on the transverse momenta of four virtual  photons. The full expression    
for $\hat{\sigma}$ is rather complicated.  However in the limit $E_t \gg    
q_{1t},q_{2t},Q_t,m_l,$ we can integrate over the direction of the lepton transverse    
momentum $\mbox{\boldmath $k$}_t$ (with $|\mbox{\boldmath $k$}_t| \simeq E_t$), and    
the formula reduces to the simple form   
\be   
\label{eq:F9}   
\hat{\sigma} \propto \frac{{\rm cosh}(\Delta\eta)}{E^4_t {\rm cosh}^4(\frac{1}{2}    
\Delta\eta)} \ \    
\frac{\left[ (Q+q_2,q_1)(Q+q_1,q_2) + (Q+q_1,q_1)(Q+q_2,q_2) -    
(q_1,q_2)(Q+q_1,Q+q_2)\right]}   
{q^2_1q^2_2(Q+q_1)^2(Q+q_2)^2}   
\ee   
where the notation $(k_1,k_2)$ is used for the scalar product of two {\it transverse} vectors,   
and where the rapidity difference    
\be   
\label{eq:F10}   
\Delta\eta \equiv \left | \eta(l^+) - \eta(l^-) \right |.   
\ee   
   
It is interesting to note that the dependence of the cross section $\hat{\sigma}$ on    
$q_{1t},~q_{2t}$ and $Q_t$ follows directly from the symmetry properties of the process.     
First, gauge invariance implies that $\hat{\sigma}$ must vanish when the transverse    
momentum of any photon goes to zero.  That is $\hat{\sigma}$ must contain a factor $q_{1t}    
q_{2t} (Q + q_1)_t (Q + q_2)_t$.  Next, it has to be symmetric under the interchanges $q_1    
\leftrightarrow q_2$ and $q_i \leftrightarrow (Q + q_i)$.  Finally, as discussed above, in the    
limit $Q_t \gg q_{1t}, q_{2t}$, it must vanish after the azimuthal angular integration of    
$\mbox{\boldmath $Q$}_t$ has been performed.  The only possibility to satisfy these    
conditions, to lowest order in $Q_t/E_t$ and $q_{it}/E_t$, is given by the expression in    
square brackets in (\ref{eq:F9}).  To obtain the $\Delta \eta$ and $E_t$ behaviour, shown in    
the first factor in $\hat{\sigma}$ of (\ref{eq:F9}), it is sufficient to put $Q_t = 0$ and to    
recall the well known pure QED cross section\footnote{We thank A.G. Shuvaev for using    
REDUCE to explicitly check form (\ref{eq:F9}) for $\hat{\sigma}$.}.   
   
Recall that for $\mu^+\mu^-$ production we select events with small transverse momentum    
of the    
lepton pair, $p_t$. Now if $\mbox{\boldmath $p$}_t = \mbox{\boldmath $q$}_{1t} -    
\mbox{\boldmath $q$}_{2t} \simeq 0$, that is if $Q_t    
\gg p_t$,    
then the last factor in expression (\ref{eq:F9}) of the absorptive cross section simplifies to    
\be   
\label{eq:F11}   
\hat{\sigma} \ \propto \ \frac{\left[ 2(Q+q_1,q_1)^2 - q^2_1(Q+q_1)^2 \right]}{\left[    
q^2_1(Q+q_1)^2 \right]^2} ,   
\ee   
which, upon integration over the azimuthal angle of the vector $(\mbox{\boldmath $Q$} +    
\mbox{\boldmath $q$}_1)_t$, gives zero.   
Therefore the main contribution to (\ref{eq:F8}) comes from values of $Q_t \approx p_t$,    
and the absorptive    
correction   
\be   
\label{eq:F12}   
\delta \approx \frac{\sigma_{\rm inel}}{8\pi} \ p^2_t \ C.   
\ee   
The coefficient $C$ is numerically small for two reasons. First, the integration over the loop    
momentum $Q_t$ kills the logarithmic factor, ln$(p^4_t/t_1t_2)$, which enhanced the    
original cross section (\ref{eq:F7}) in the absence of absorptive corrections.  To see this, note    
that the $\log$ came from the $q_t^2 dq_t^2/q_t^4$ integrations, for both $q_{1t}$ and    
$q_{2t}$, of (\ref{eq:F4}), see (\ref{eq:F7}).  However when $Q_t \neq 0$ the integrands are    
of the form $\mbox{\boldmath $q$}_t \cdot (\mbox{\boldmath $Q$}_t + \mbox{\boldmath    
$q$}_t)/q_t^2 |\mbox{\boldmath $Q$}_t + \mbox{\boldmath $q$}_t |^2$, and there is no    
logarithmic singularity for either $q_t \rightarrow 0$ or $|\mbox{\boldmath $Q$}_t +    
\mbox{\boldmath $q$}_t| \rightarrow 0$.  Second, there is a suppression of $C$ from    
cancellations which occur after integration over the remaining azimuthal angles.  For    
example, if the invariant mass of the $\mu^+ \mu^-$ pair produced at zero rapidity is    
$M_{\mu\mu} = 20~{\rm GeV}$, then the value of the coefficient $C = 0.14, 0.13, 0.09$ and   
0.08 for $p_t = 5, 10, 30$ and 50~MeV respectively.  This leads to a negligible correction to   
the cross section (\ref{eq:F7}), for example   
\be   
\label{eq:F13}   
2 \delta \lapproxeq \frac{80{\rm mb}}{4 \pi} \ p^2_t \ C < 0.02\%~(0.13\%)   
\ee   
for $p_t = 10$ MeV (30~MeV).   
  
Note that the rescattering contribution of Fig.~3(b) is less singular as $q_{1t}, q_{2t}   
\rightarrow 0$ than the pure QED term of Fig.~3(a).  Therefore the rescattering correction   
does not induce a sharp peak at $\phi = 0$ in the muon acoplanarity distribution.  Thus   
removing the excitation processes (\ref{eq:F1}) by fitting the $\phi$ distribution, will   
automatically suppress the rescattering correction.  
   
\subsection{$e^+e^-$ production}   
   
Unlike $\mu^+ \mu^-$ production, for $e^+ e^-$ production we do not need to select events    
with electrons of large transverse momentum $p_{et}$.  Providing the energy of the electrons    
$E \sim 5~{\rm GeV}$, we may use, say, an electromagnetic calorimeter to identify them.     
Therefore we may consider the small $p_{et}$ domain where the $e^+ e^-$ production cross    
section is much larger.  The dominant contribution to the cross section comes from the region    
of $p_{et}$ of the order of the electron mass ($p_{et} \lapproxeq 1$ MeV).  As $p_{et}$ is    
small, we cannot neglect $Q_t$ in the $t$ channel electron propogator. As a consequence the    
integral over $Q_t$ becomes super-convergent   
\be   
\label{eq:F14}   
\int \frac{d^2Q_t |\mbox{\boldmath $Q$}_t + \mbox{\boldmath $q$}_{1t}| \:    
|\mbox{\boldmath $Q$}_t + \mbox{\boldmath $q$}_{2t}|}{(Q + q_{1})^2 \: (Q + q_{2})^2    
\: (|Q + q_1 + p_e|^2 + m^2_e)}.   
\ee   
It is informative to note the origin of this form.  Due to gauge invariance, the numerator    
vanishes when the photon transverse momenta $|\mbox{\boldmath $Q$}_t +    
\mbox{\boldmath $q$}_{it}| \rightarrow 0$.  The first two factors in the denominator arise    
from the photon propagators.  Finally we have the factor due to the electron propagator, with    
$Q \simeq (0; \mbox{\boldmath $Q$}_t, 0)$.  Contrary to $\mu^+ \mu^-$ production, where    
the lepton propagator was driven by the large $p_t \simeq E_t$ of the muon, here the    
momentum $Q_t$ is not negigible in comparison with $p_{et}$ and the dominant    
contribution comes from the region $Q^2_t \simeq m^2_e < 10^{-6}$~GeV$^2$. Hence the    
expected absorptive correction due to strong interaction rescattering is   
\be   
\label{eq:F15}   
2 \delta \approx \frac{\sigma_{\rm inel}}{4\pi} \ Q^2_t \sim \frac{80{\rm mb}}{4\pi} \ 10^{-   
6} {\rm GeV}^2 \lapproxeq 10^{-5}   
\ee   
   
\subsection{Conclusion on $l^+l^-$ production as a luminosity monitor}   
   
We see that both $\mu^+\mu^-$ and $e^+e^-$ pair production may be used as monitor    
processes in high luminosity LHC collisions. However we must employ special selection cuts    
on the $l^+l^-$ data. To suppress $N^*$ contamination we impose small transverse    
momentum $p_t$ of the lepton pair, moreover for $e^+e^-$ production we require the    
individual electron $p_{et}$ to be small. We also require the leptons to be sufficiently    
energetic in order to identify them. Then the cross section for $l^+l^-$ production at the LHC    
$(pp \rightarrow p + l^+l^- + p)$ can be calculated within pure QED and, {\it importantly},    
we may neglect the strong rescattering effects between the protons up to $10^{-4}$ accuracy.    
Of course by requiring the muons to have high $E_t$ we have a rather small cross section.    
However here we may trace back the muon tracks and determine the interaction vertex, and    
hence isolate the interaction in pile-up events. So, in principle, $\mu^+\mu^-$ pair    
production, with high $E_t$ muons, may be used as a luminometer in very high luminosity   
LHC runs.   
   
\section{$W$ and $Z$ production as a luminosity monitor}   
   
$W$ and $Z$ production in high energy $pp$ and $p\bar{p}$ collisions have clean signatures    
through their leptonic decay modes, $W \rightarrow l\nu$ and $Z \rightarrow l^+l^-$, and so    
may be considered as potential luminosity monitors \cite{R12}. A vital ingredient is the    
accuracy to which the cross sections for $W$ and $Z$  production can be theoretically    
calculated. The cross sections depend on parton distributions, especially the quark densities,    
in a kinematic region where they are believed to be reliably known. Recent determinations of    
the $W$ and $Z$ cross sections can be found in Refs. \cite{MRST1,MRST}. Here the    
situation has improved, and next-to-next-to-leading order (NNLO) predictions have been    
made\footnote{For a precise measurement, we should allow for $W^+ W^-$ pair production   
and for $W$ bosons produced via $t$-quark decays.  These can contribute about 1\% of the   
total signal.}. The most up-to-date values are reproduced in Fig.~4 \cite{MRST} and   
Figs.~5,6 \cite{MRST1,MRST}.   
   
To estimate the accuracy with which the $W$ and $Z$ cross sections are known, we start    
with Fig. 4.   Fig. 4 was obtained using parton distributions found in LO, NLO and NNLO    
global analyses of the same data set \cite{MRST}. It is relevant to summarize the contents of    
these plots. The predictions labelled LO, NLO and NNLO can be schematically written as    
follows   
\bea      
\label{eq:F16}    
\sigma_{\rm LO} &=& f_{\rm LO} \otimes f_{\rm LO} \nonumber\\   
\sigma_{\rm NLO} &=& f_{\rm NLO} \otimes f_{\rm NLO} \otimes [1 + \alpha_{S,{\rm    
NLO}} K^{(1)}] \nonumber\\   
\sigma_{\rm NNLO} &=& f_{\rm NNLO} \otimes f_{\rm NNLO} \otimes [1 +    
\alpha_{S,{\rm NNLO}} K^{(1)} + (\alpha_{S,{\rm    
NNLO}})^2 K^{(2)}]   
\eea   
where the label on $\alpha_S$ indicates the order to which the $\beta$-function is evaluated.    
The parton distributions $f$ obtained from the new global LO, NLO and NNLO analyses    
\cite{MRST} correspond to $\alpha_S(M^2_Z) =$ 0.1253, 0.1175 and 0.1161 respectively.    
The NLO and NNLO coefficient functions $K^{1,2}$ are known \cite{HMVN}. However,    
although the relevant deep inelastic coefficient functions are also known, there is only partial    
information on the NNLO splitting functions. Recently van Neerven and Vogt \cite{VN}    
have constructed compact analytic expressions for the NNLO splitting functions which   
represent the fastest and the slowest possible evolution that is consistent with the existing    
partial information. These expressions have been used to perform global parton analyses at    
NNLO \cite{MRST}.  The uncertainty in the NNLO predictions of $\sigma_W$ and    
$\sigma_Z$ due to the residual ambiguity in the splitting functions is shown by the width of    
the NNLO bands in Fig.~4. This amounts to about $\pm 1\%$ uncertainty at the LHC energy,    
and less at the Tevatron. For completeness, we note that the dashed lines in Fig.~4 correspond    
to the quasi-NLO prediction    
\be      
\label{eq:F17}    
\sigma_{{\rm NLO}^\prime} = f_{\rm NLO} \otimes f_{\rm NLO} \otimes [1 +    
\alpha_{S,{\rm NLO}} K^{(1)} + (\alpha_{S,{\rm    
NLO}})^2 K^{(2)}] ,   
\ee   
which was the best that could be done before the work of Refs. \cite{VN,MRST}.   
   
We see that the LO $\rightarrow$ NLO $\rightarrow$ NNLO convergence of the predictions    
for $\sigma_{W,Z}$ is good. The jump from $\sigma_{\rm LO}$ to $\sigma_{\rm NLO}$ is    
mainly due to the well-known, large, $O(\alpha_S)$ $\pi^2$-enhanced Drell-Yan K-factor    
correction, arising from soft-gluon emission. The NLO and NNLO cross sections are much    
closer and, if this was the end of the story, $W$ and $Z$ production can clearly be predicted    
with sufficient accuracy.     
   
However let us turn to Figs.~5 and 6, each of which combine results presented in Refs.    
\cite{MRST1} and \cite{MRST}.  The solid squares and triangles show the additional   
uncertainty in the predictions for $\sigma_W$ and $\sigma_Z$ which arise from changing the   
input information in the global parton analyses. The two major uncertainties appear to be due   
to the value of $\alpha_S$ and to using different parton densities labelled by q$\uparrow$ and    
q$\downarrow$. The plots show the change in $\sigma_W$ and $\sigma_Z$ which is caused    
by changing the value of $\alpha_S(M^2_Z)$ by $\pm0.005$ respectively.  The change in    
$\sigma_W$ and $\sigma_Z$ at the LHC energy is enhanced as compared to that at the    
Tevatron, since DGLAP evolution is more rapid at the smaller $x$ values, $x \sim M_{W,Z}    
/ \sqrt{s}$, probed at the higher energy. However, with our present knowledge of $\alpha_S$,    
the uncertainty $\Delta\alpha_S = \pm 0.005$ is too conservative, and $\pm 2\%$ is a more    
realistic uncertainty in $\sigma_{W,Z}$ from this source at the LHC energy.   
   
The normalisation of the input data used in the global parton analyses is another source of    
uncertainty in  $\sigma_{W,Z}$. The HERA experiments provide almost all of the data used    
in the global analyses in the relevant small $x$ domain.  The quoted normalisation    
uncertainties of the measurements of the proton structure function $F_2$ from the H1 and    
ZEUS experiments vary with $Q^2$, but a mean value of $\pm 2.5 \%$ is appropriate. The    
q$\uparrow$ and q$\downarrow$ parton sets correspond to separate global fits in which the    
HERA data have been renormalised by $\pm 2.5 \%$ respectively. For $W$ and $Z$    
production, of $q\bar{q}$ origin, we naively would expect this to translate    
into a $\pm 5 \%$ variation in $\sigma_{W,Z}$, but the effect of DGLAP evolution up to    
$Q^2 \sim M^2_{W,Z}$ is to suppress the difference in the predictions.   
   
In summary, allowing for all the above uncertainties, we conclude that the cross sections of    
$W$ and $Z$ production are known to $\pm 4 \%$ at the LHC energy, and to $\pm 3 \%$ at    
the Tevatron. A major contributer to this error is the uncertainty in the overall normalisation    
of the H1 and ZEUS measurements of $F_2$.  The normalisation may be made more precise    
by experiments in Run II at the Tevatron.   
   
\section{Parton-parton luminosity}    
    
In some circumstances it is sufficient to know the parton-parton luminosity, and not the     
proton-proton luminosity, see, for example, \cite{R12}.  Of course if the proton-proton    
luminosity is known     
then the parton-parton luminosities can be calculated from the parton distributions determined     
in the global parton analyses.  However in this case we rely on the normalisation of     
experiments at previous accelerators which yielded data that were used in the global analyses.    
    
Thus it may be better to monitor the parton-parton luminosities directly in terms of a     
subprocess which can be predicted theoretically to high precision.  The best example is     
inclusive $W$ (or $Z$) boson production, which is predicted up to two-loops, that is to    
NNLO \cite{MRST}.  The accurate observation of $W$ (or $Z$) production at the LHC may     
therefore be used to determine the quark-antiquark luminosity\footnote{At first sight the    
$q\bar{q}$ luminosity in a given $x_1, x_2$ bin may be obtained by observing the number of    
$W$ events in that bin and dividing by the $q\bar{q} \rightarrow W$ cross section.  However    
at NLO we include $q\bar{q} \rightarrow Wg, qg \rightarrow Wq$ etc., so the only possibility    
is to use the new $W$ data to determine the $q\bar{q}$ luminosity within a global parton    
analysis.  To do the same at NNLO we would require the NNLO expression for    
$d\sigma/dy_\ell$, where $y_\ell$ is the rapidity of the decay lepton.}.  Then the gluon flux,    
for example, may be determined from the global parton analysis which already is made to     
describe the measured $W$ (or $Z$) cross sections.  One advantage of this technique is that,     
for most LHC applications such as the search for new heavy particles, we need to consider a    
smaller interval of DGLAP evolution than has been the practice hitherto.    
    
Other ways to constrain the gluon-gluon luminosity are to study $t\bar{t}$ production or the   
production of two large $p_t$ prompt photons.  The leading order subprocess is $q\bar{q}   
\rightarrow \gamma\gamma$.  However at LHC energies the $gg \rightarrow   
\gamma\gamma$ box diagram gives an important contribution.  The two diagrams are shown   
in Fig.~7.  The relative contributions are shown in Fig.~8 as a function of the transverse   
momentum of the photon pair $\mbox{\boldmath $p$}_t = \mbox{\boldmath $q$}_{\gamma   
1 t} + \mbox{\boldmath $q$}_{\gamma 2 t}$ for the case when each photon has transverse    
momentum $q_{\gamma t} > 20~{\rm GeV}$ and rapidity $| \eta_\gamma | < 1$.  We see    
that in this kinematic domain the $gg \rightarrow \gamma\gamma$ subprocess gives a major    
contribution due to the higher $gg$ luminosity.  However there is a strong possibility of    
contamination by the subprocess $gq \rightarrow \gamma\gamma q$, unless severe photon    
isolation cuts are imposed, see, for example, Ref.~\cite{R1}.   
    
So far we have considered conventional parton distributions $a (x, \mu^2)$ integrated over     
the parton transverse momentum $k_t$ up to the factorization scale $\mu$.  However many     
reactions are described by unintegrated distributions $f_a (x, k_t^2, \mu^2)$ which depend on     
both $k_t$ and the longitudinal momentum fraction $x$ carried by the parton.  In principle,    
unintegrated distributions are necessary for the description of all processes which are not    
totally inclusive.  In these cases, instead of the conventional QCD factorization, we have    
$k_t$ factorization \cite{KTFAC}.  Indeed, for some processes, after the specific integration    
over $k_t$, the conventional \lq hard\rq\ QCD factorization may be destroyed.    
    
The unintegrated quark-quark luminosity may be determined, for example, by observing the    
$k_t$ distribution of $W$ (or $Z$) bosons or Drell-Yan pairs.  At leading order, the     
$k_t$-dependence of cross sections is given by the convolution   
\be    
\label{eq:2e}    
\frac{d\sigma}{d^2 k_t} \; \propto \; \int \: \frac{d^2 p_1}{p_1^2} \: \frac{d^2 p_2}{p_2^2}    
\: \delta^{(2)} (\mbox{\boldmath $p$}_1 + \mbox{\boldmath $p$}_2 - \mbox{\boldmath    
$k$}_t) \: f_q (x_1, p_1^2, \mu^2) \: f_{\bar{q}} (x_2, p_2^2, \mu^2).   
\ee    
At large $k_t \gg \Lambda_{\rm QCD}$, to leading $\ln k_t$ order, the dominant    
contribution comes from either the domain $p_1 \approx k_t,~p_2 \ll k_t$ or the domain    
$p_2 \approx k_t,~p_1 \ll k_t$.  It is natural to choose the factorization scale $\mu \approx    
k_t$.  Then the $k_t$    
distribution takes the simple form \cite{DDT}   
\be   
\label{eq:2f}   
\frac{d\sigma}{d^2 k_t} \; \propto \; \left [x_1 q_1 (x_1, k_t^2) \: f_{\bar{q}} (x_2, k_t^2,    
k_t^2) \: + \: x_2 \bar{q}_2 (x_2, k_t^2) \: f_q (x_1, k_t^2, k_t^2) \right ],   
\ee   
since at leading order   
\be   
\label{eq:2g}   
\int^{k_t^2} \: \frac{dp^2}{p^2} \: f_a (x, p^2, k_t^2) \; = \; xa (x, k_t^2).   
\ee   
Moreover diffractive processes (with one or more rapidity gaps) are described, in general, by     
skewed parton distributions.  An example is the double-diffractive exclusive Higgs boson     
production, described by the Feynman diagram shown in Fig.~9.  This process is described by     
skewed gluon distributions with $x_1 \neq x_1^\prime$ and $x_2 \neq x_2^\prime$ 
\cite{KMR1,KMR2}.    
    
Sometimes even the \lq\lq skewed\rq\rq\ parton flux can be monitored directly via another     
process with similar kinematics.  For example, we may monitor the effective     
Pomeron-Pomeron luminosity for Higgs production of Fig.~9 by measuring double-    
diffractive dijet production in the region in which the transverse energy of the jets ($E_t$) is     
of about half that of the Higgs mass $(E_t \sim M_H/2)$ \cite{KMR2,KMR3}.    
    
Fortunately most of the physically relevant processes are described by skewed distributions     
with $x_i \ll 1$.  In these cases the skewed distributions may be reliably reconstructed from     
the known conventional parton distributions \cite{SGMR}.    
   
\section{Conclusions}   
   
We have studied the theoretical accuracy of the main three proposals for determining the    
luminosity of the LHC $pp$ collider --- namely using forward elastic data, lepton-pair    
production and $W$ or $Z$ boson production.  The desired goal of a measurement to within    
$\pm 5\%$ seems theoretically attainable.   
   
We focused on potential shortcomings of each method.  First, we demonstrated that the $t$    
dependence of the elastic cross section is well under control and, in fact, it turns out that it    
can be safely approximated by a simple exponential in the region $|t| < 0.05~{\rm GeV}^2$.     
For lepton-pair production we evaluated the corrections to the cross section arising from the    
strong interactions between the protons.  We showed that in the relevant kinematic domain,    
with small transverse momentum of the produced lepton-pair, these effects are negligible.     
Hence a pure QED calculation of the cross section will give sufficient accuracy.   
   
The cross sections for $W$ and $Z$ production can now be predicted to NNLO, which at first    
sight would seem to provide an LHC luminometer with $\pm 1\%$ accuracy, see Fig.~4.     
However the uncertainties in the input to the global parton analyses mean that the error could,   
conservatively, be as large as $\pm 4\%$.  The uncertainty may be reduced if we work    
in terms of quark-antiquark luminosity, which is relevant in some future applications of the    
LHC.   
  
Finally, we note that luminosity determinations based on the measurement of the forward   
elastic cross section and (most probably) on two-photon $e^+ e^-$ production can only be   
made in low luminosity runs, and require dedicated detectors and triggers.  On the other   
hand, the measurement of $W$ or $Z$ and two-photon $\mu^+ \mu^-$ production may be   
performed at high luminosity with the central detector and with standard triggers.  
   
\section*{Acknowledgements}   
   
We thank V.S. Fadin, M.A. Kimber, V. Nomokonov, A. Shamov, A.G. Shuvaev, W.J.    
Stirling and S. Tapprogge for useful discussions.  VAK thanks the Leverhulme Trust for a    
Fellowship.  MGR thanks the Royal Society and PPARC for support, and also the Russian    
Fund for Fundamental Research (98-02-17629).  This work was also supported by the EU    
Framework TMR programme, contract FMRX-CT98-0194 (DG 12-MIHT).

\newpage   

\begin{figure}[t]
\begin{center}
\epsfig{figure=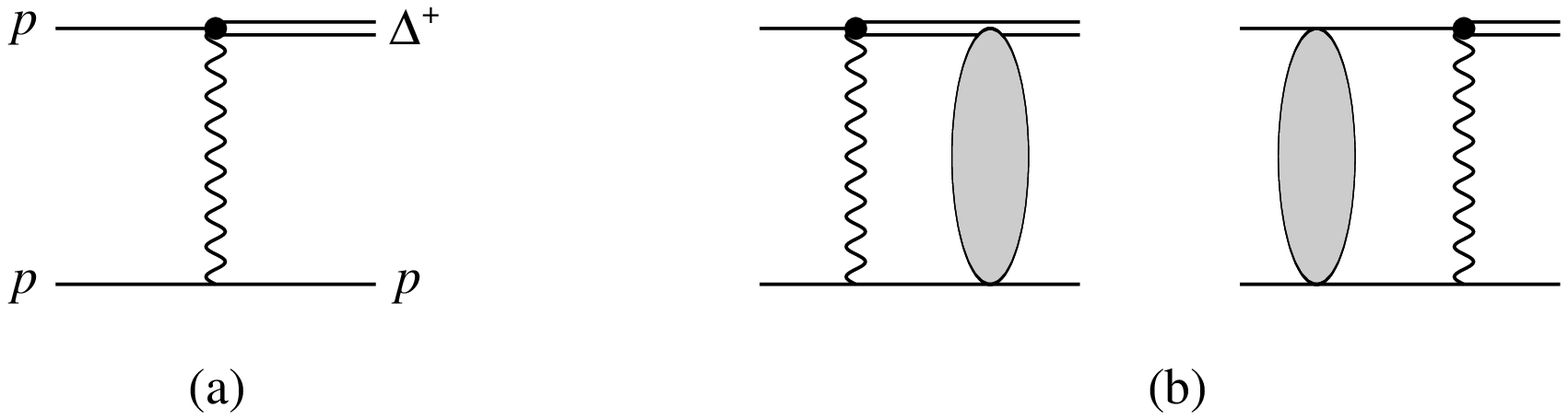}
\end{center}
\caption{
(a) $\Delta^+$ production mediated by photon exchange, and (b) possible    
rescattering corrections.
}
\end{figure}

\begin{figure}[t]
\begin{center}
\epsfig{figure=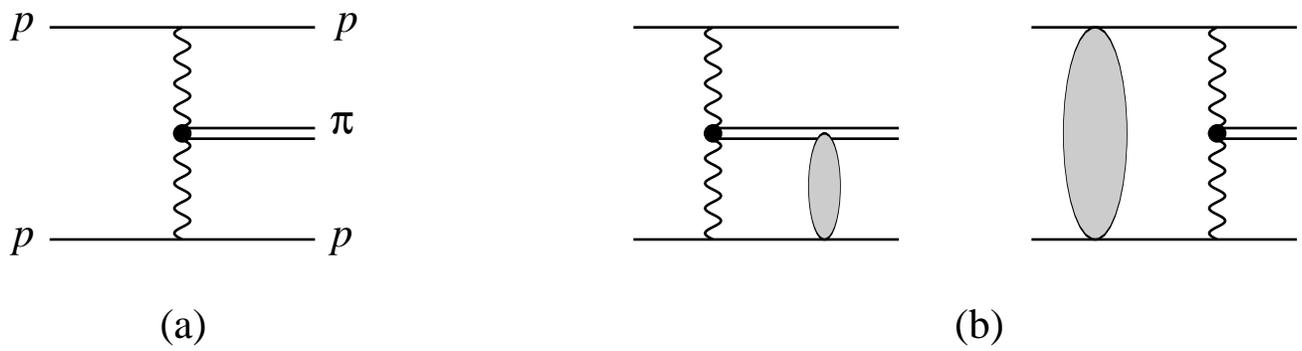}
\end{center}
\caption{
(a) $\pi^0$ production mediated by photon exchanges and (b) two of the   
possible rescattering corrections. 
}
\end{figure}

\begin{figure}[t]
\begin{center}
\epsfig{figure=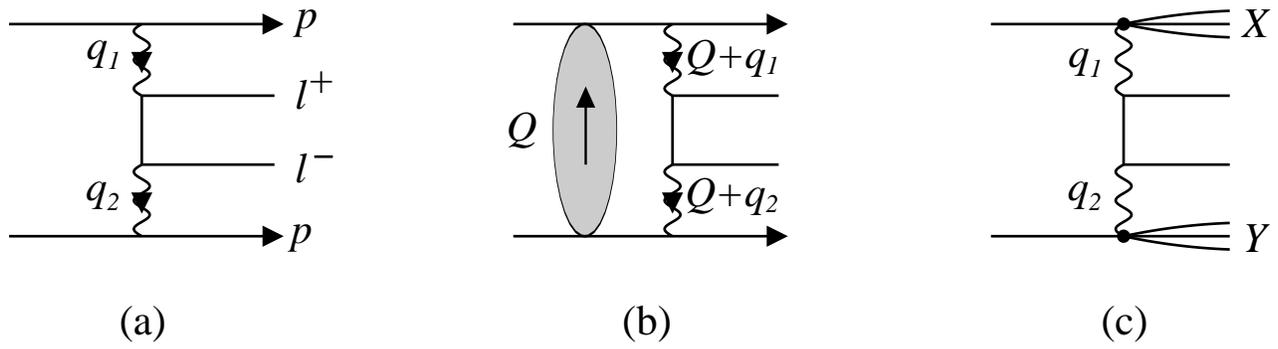}
\end{center}
\caption{
(a) Lepton pair production in $pp$ collisions, (b) one of the rescattering    
corrections, and (c) a possible contamination coming from proton dissociation into $X, Y$  
systems.   
}
\end{figure}

\begin{figure}[t]
\vspace*{-1.0cm}
\begin{center}
\epsfig{figure=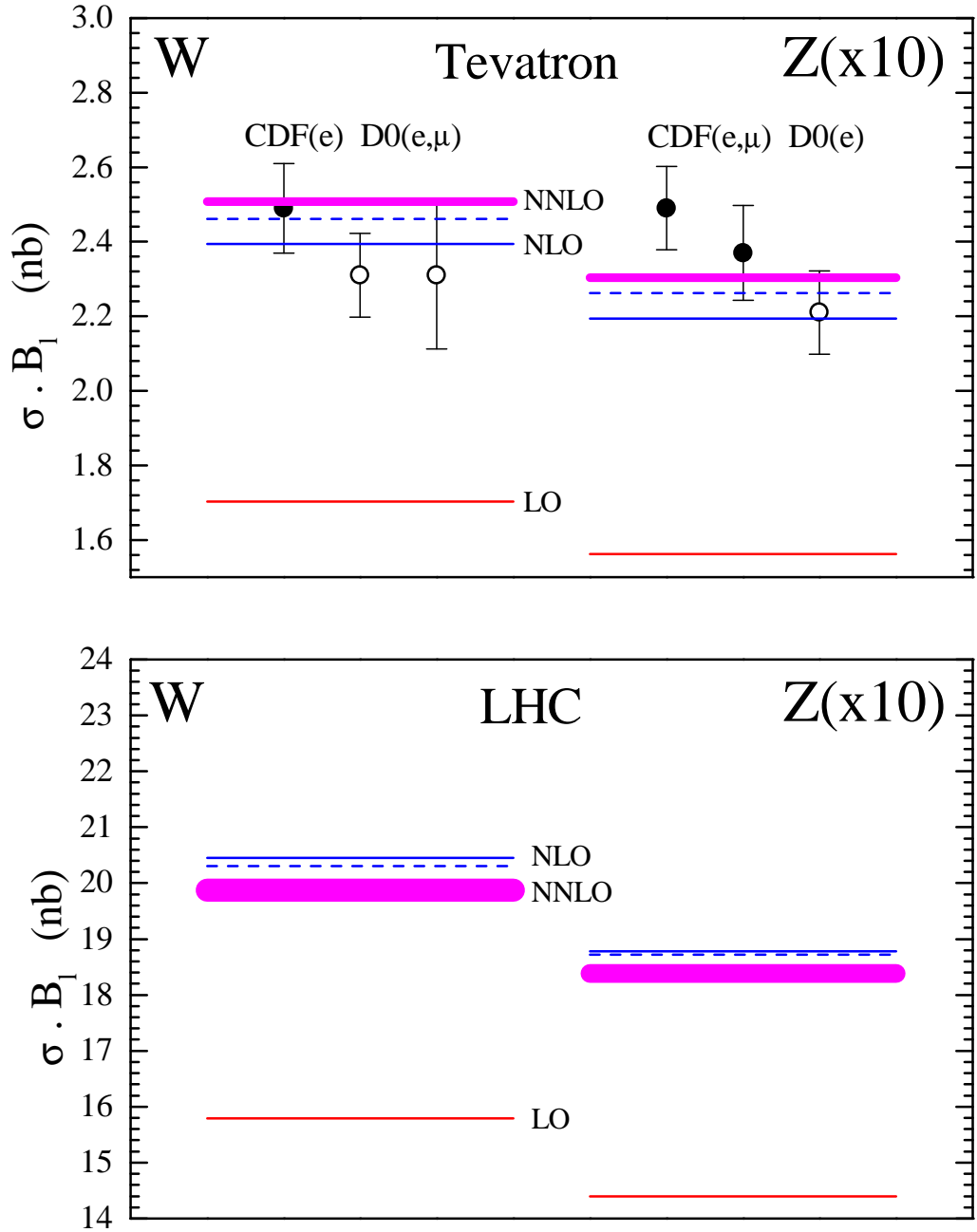,height=18cm}
\end{center}
\caption{
The predictions of the cross sections for $W$ and $Z$ production and leptonic    
decay at the Tevatron and the LHC obtained from parton sets of LO, NLO and NNLO global    
analyses of the same data set \cite{MRST}.  The cross sections labelled LO, NLO and NNLO    
are as in (\ref{eq:F16}) and the dashed line is the NLO$^\prime$ prediction of (\ref{eq:F17}).     
The band of the NNLO predictions allow for the ambiguity in the NNLO splitting functions    
\cite{VN}.  Also shown are measurements obtained at the Tevatron \cite{WZCDF,WZD0}.     
The figure is taken from \cite{MRST}. 
}
\end{figure}

\begin{figure}[t]
\vspace*{-1.0cm}
\begin{center}
\epsfig{figure=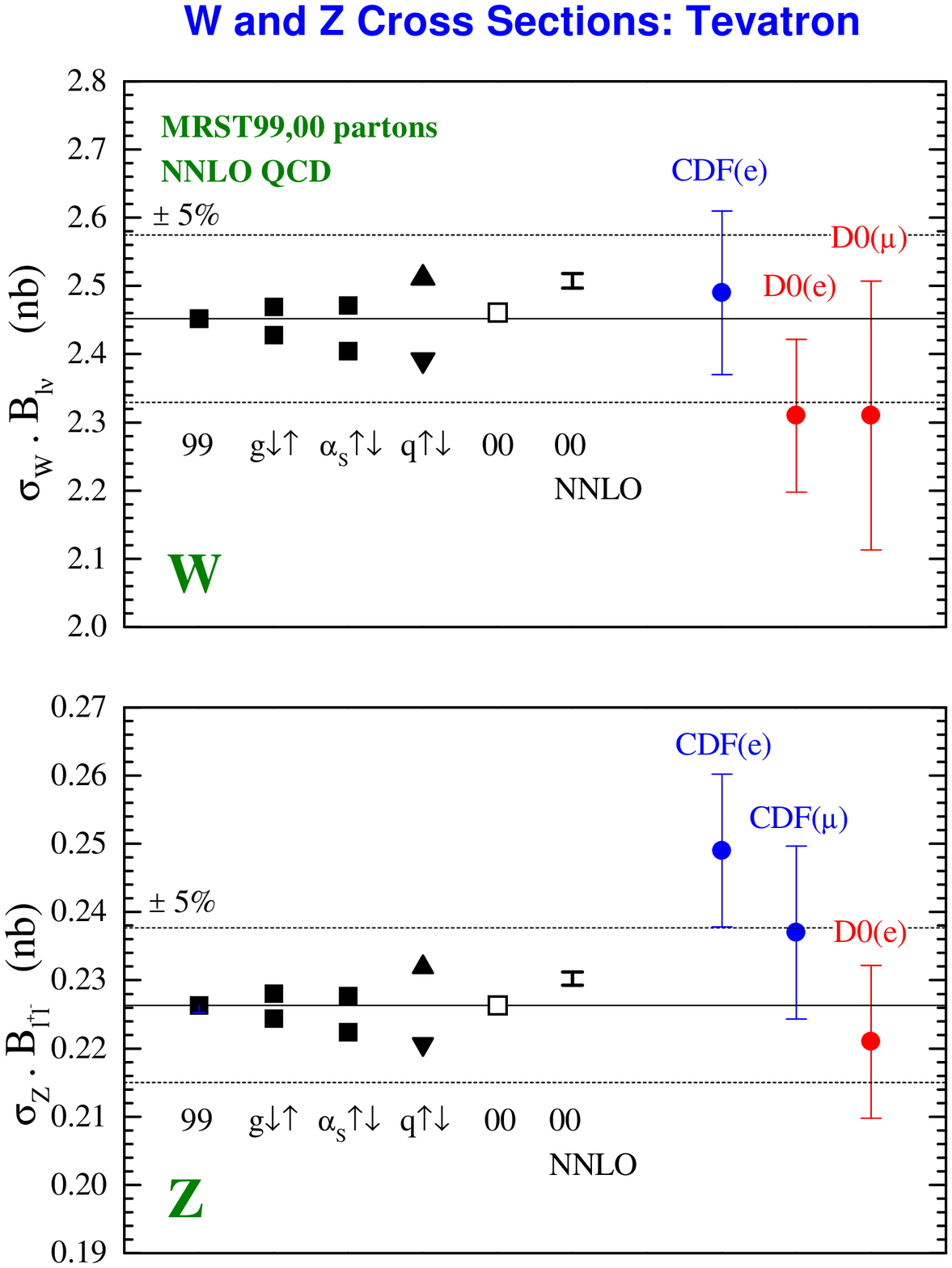,height=18cm}
\end{center}
\caption{
The solid squares and triangles are the predictions of the NLO$^\prime$ cross    
sections of (\ref{eq:F17}) for $W$ and $Z$ production and leptonic decay in $p\bar{p}$    
collisions at $\sqrt{s} = 1.8~{\rm TeV}$ obtained using various NLO sets of MRST99    
partons \cite{MRST1}.  The open square and small error bar are, respectively, the NLO and    
NNLO predictions of (\ref{eq:F16}) using the MRST00 partons \cite{MRST}.  Also shown    
are the experimental measurements from CDF \cite{WZCDF} and D0 \cite{WZD0}.  For    
ease of reference $\pm 5\%$ lines are shown about the MRST99 default prediction.  We    
thank W.J. Stirling for this figure, which combines results presented in   
\cite{MRST1,MRST}. 
}
\end{figure}

\begin{figure}[t]
\vspace*{-1.0cm}
\begin{center}
\epsfig{figure=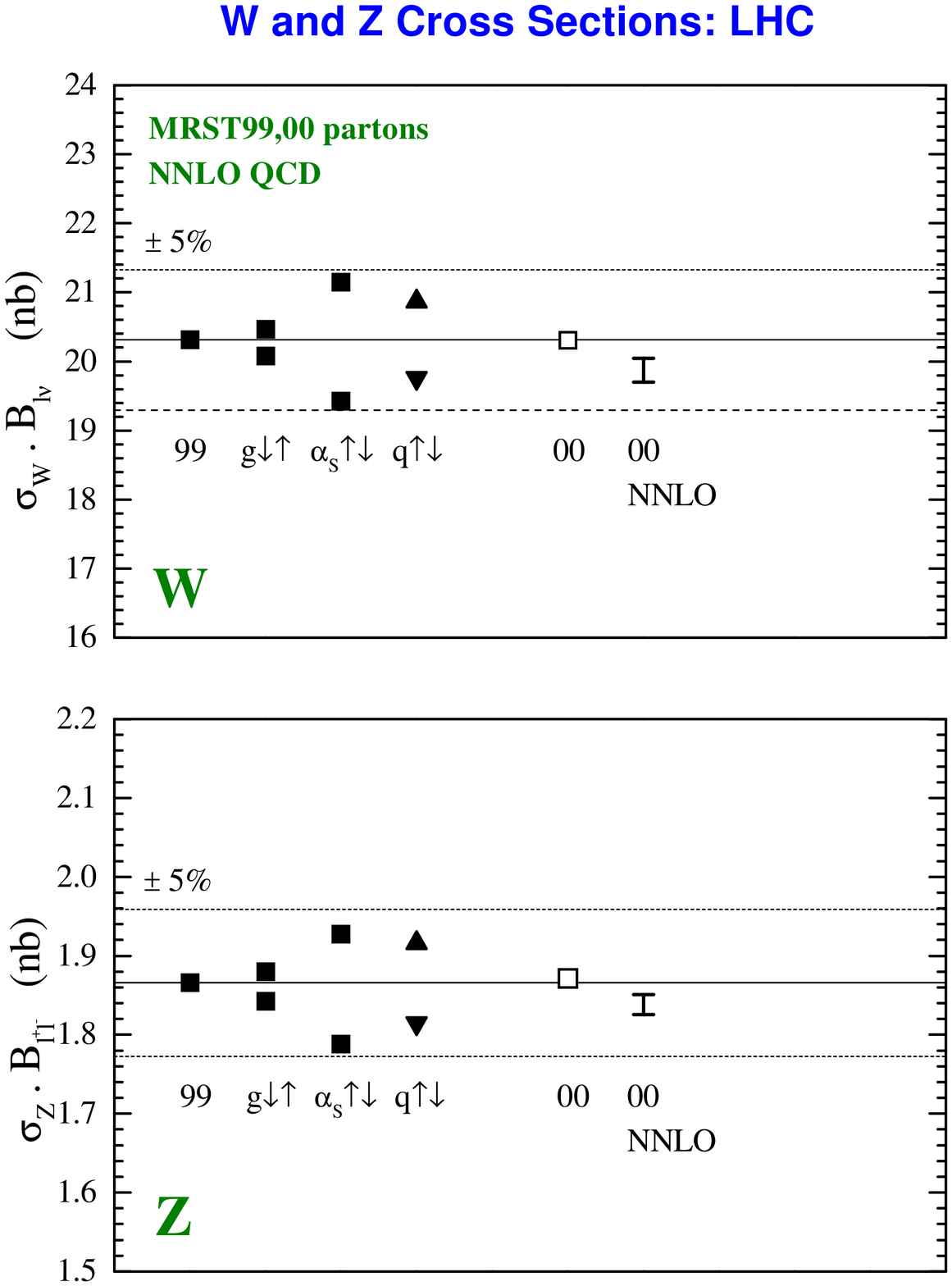,height=18cm}
\end{center}
\caption{
As for Fig. 5 but for $pp$ collisions at $\sqrt{s} = 14~{\rm TeV}$. 
}
\end{figure}

\begin{figure}[t]
\begin{center}
\epsfig{figure=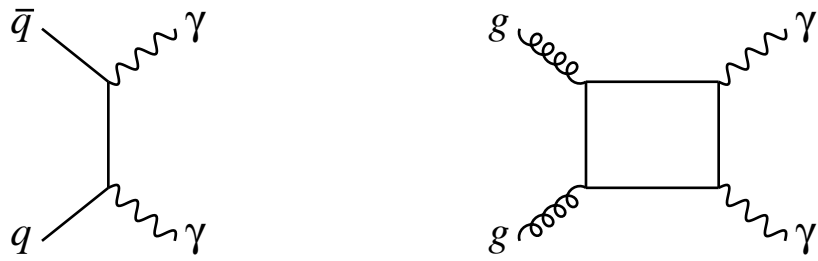}
\end{center}
\caption{
Feynman diagrams driving $\gamma\gamma$ production in $pp$ collisions.   
}
\end{figure}

\begin{figure}[t]
\begin{center}
\epsfig{figure=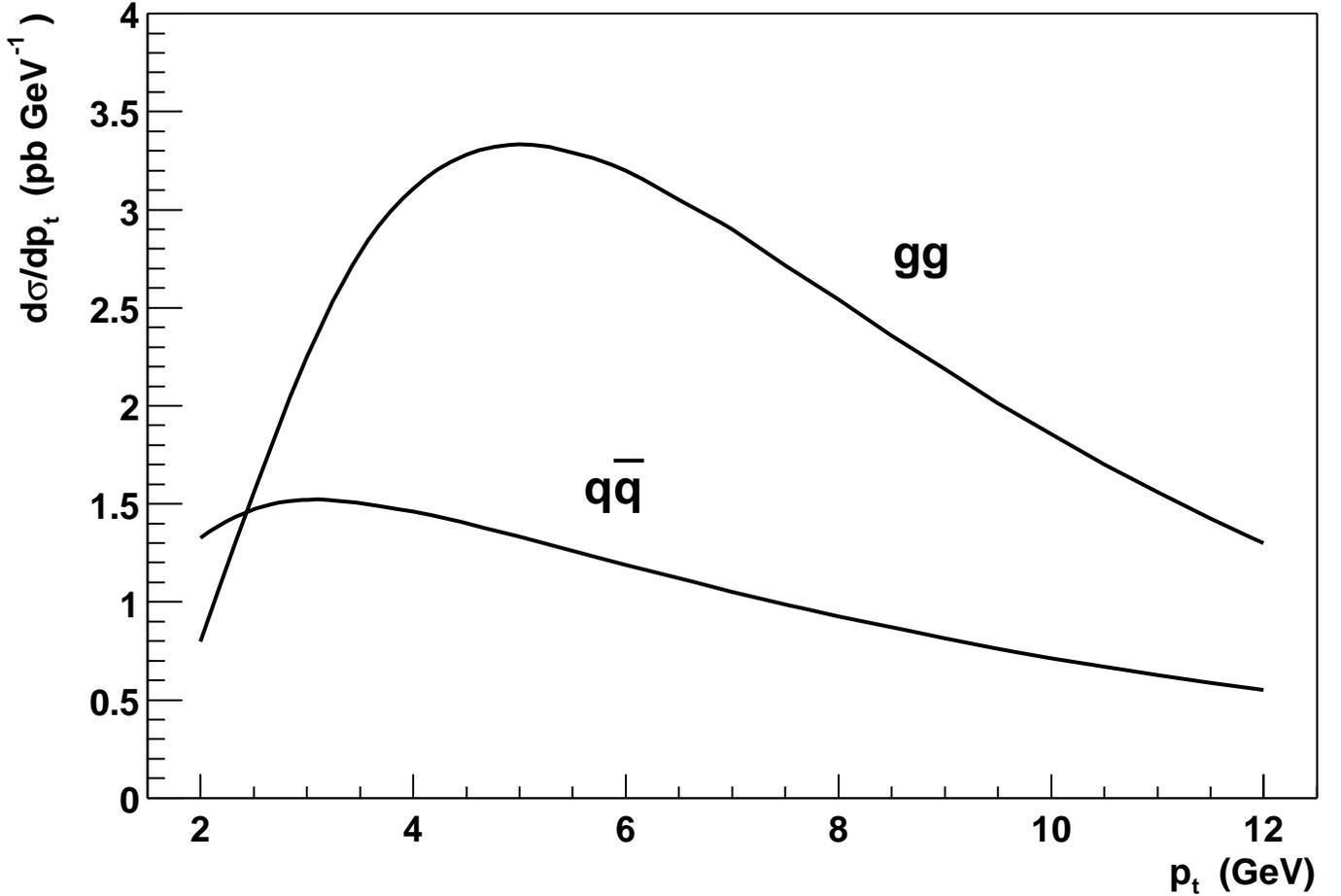}
\end{center}
\caption{
Contributions to the diphoton $p_t$ spectrum from the diagrams of Fig. 7 in    
$pp$ collisions at $\sqrt{s} = 14~{\rm TeV}$.  Each photon is required to have transverse    
momentum $p_{\gamma t} > 20~{\rm GeV}$ and rapidity $| \eta_\gamma | < 1$.  The    
photons are required not to lie within the same $\eta - \phi$ cone of radius 0.4.  We thank    
M.A. Kimber for this figure.   
}
\end{figure}

\begin{figure}[t]
\begin{center}
\epsfig{figure=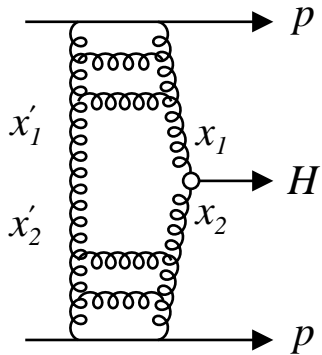}
\end{center}
\caption{
Higgs production via Pomeron-Pomeron fusion in $pp$ collisions.  
}
\end{figure}

\end{document}